\documentclass[conference]{IEEEtran}
\usepackage{graphicx}
\usepackage{epstopdf} 
\usepackage{verbatim} 
\usepackage{multirow} 
\usepackage{multicol}
\usepackage{amsfonts}
\usepackage{bbm}
\usepackage{subfigure}
\usepackage{colortbl} 
\usepackage{indentfirst}
\usepackage{dcolumn,booktabs}
\usepackage{algorithm}
\usepackage[noend]{algpseudocode}
\usepackage{algorithmicx,algorithm}
\usepackage{setspace}

\usepackage{algpseudocode}
\usepackage{graphics}
\usepackage{epsfig,color}
\usepackage{amsmath}
\usepackage{amssymb}

\usepackage{color}

\ifCLASSINFOpdf

\else
\fi
\newtheorem{Theorem}{Theorem}
\newtheorem{Corollary}{Corollary}

\newtheorem{Lemma}{Lemma}

\newtheorem{Remark}{Remark}

\begin{document}
\begin{spacing}{1.0}
\title{Coverage Probability and Energy Efficiency of Reconfigurable Intelligent Surface-Assisted mmWave Networks}

\author{
\IEEEauthorblockN{Le Yang, Fu-Chun Zheng, and Shi Jin}
}
\maketitle

\begin{abstract}
Reconfigurable intelligent surface (RIS) emerges as a promising technology for the next generation networks. In this paper, we utilize the tools from stochastic geometry to study the performance of a RIS-assisted millimeter wave (mmWave) cellular network. Specifically, the locations of the base stations (BSs) and the midpoints of the blockages are modeled as two independent Poisson point processes (PPP) where the blockages are modeled by line boolean model and a fraction of the blockages are coated with RISs. The distinguish characteristics of mmWave communications, i.e., the directional beamforming and different path loss laws for line-of-sight (LOS) and non-line-of-sight (NLOS), are incorporated into the analysis. We derive the expressions of the coverage probability and the area spectral efficiency. The coverage probability under the special case where the blockage parameter is sufficiently small is also derived. Numerical results demonstrate that better coverage performance and higher energy efficiency can be achieved by the large-scale deployment of RISs. In addition, the tradeoff between the BS and RIS densities is investigated and the results show that the RISs are excellent supplementary for the traditional networks to improve the coverage probability with limited power consumption.
\end{abstract}


%
\IEEEpeerreviewmaketitle

\section{Introduction}
Due to the rapid proliferation of various smart mobile devices and multi-media applications, the mobile data traffic has witnessed an unprecedented growth. Therefore, it is one of the main challenges to cope with the demand for dramatically increasing data rate in the next generation wireless networks. Utilizing the the higher frequency band, i.e., the millimeter wave (mmWave), is a promising method to tackle the problem since the available spectrum in the millimeter wave band is much broader than the sub-6GHz \cite{mmWave-5G}. Compared with the sub-6GHz, the drawback of the millimeter wave is the sensitivity to the blockages. Although the comparable coverage can be achieved by deploying the directional array antenna \cite{measurement-1}\cite{measurement-2}, hindrances still occur due to the existence of blockages. To address this problem, a tradition approach is to add the supplementary links to enhance the signal strength experienced by the receiver. For example, a relay can be deployed to receive the signals, then the signals are amplified and transmitted to the receiver.

In recent years, reconfigurable intelligent surface (RIS) has attracted considerable attentions from both academia and industry \cite{tang-1}\cite{tang-2}. A RIS is a surface consisting of a number of reflective elements, which can obtain preferable electromagnetic propagation environment through digitally manipulating electromagnetic waves. The RIS and the relay work in different mechanisms to assist the communication. By connected to a simple controller, the phase shifters on the RIS can be adjusted and the impinging signals can be reflected toward desired spatial direction by adjusting the phase shifters. The signal can be forwarded by the RIS without causing additional noise. In addition, no self-interference is caused when the RIS is utilized in the full-duplex mode. The authors in \cite{ris-versus-relay} compared the performance of RIS-assisted and relay-assisted networks. The results demonstrated that RIS-assisted networks may have better performance than the relay-assisted networks, providing enough number of RISs.

Since the RISs can be utilized to provide the supplementary paths for the blockage links with limited power consumption, performing a system-level analysis on the performance enhancement by introducing the RISs into the cellular networks is important. By coating the existing objects, i.e., buildings and trees, with RISs, these objects can be exploited to enhance the network performance. In this paper, we utilize tools from stochastic geometry to investigate the performance gain in terms of coverage probability, area spectral efficiency (ASE) and energy efficiency.

\subsection{Related Works}
The RISs can be utilized in various scenarios, such as interference engineering, signal strength enhancement and signal secrecy at eavesdroppers. The results in \cite{confirm} confirmed the huge potential of RISs to enhance the secrecy performance. In \cite{minimax}, the BS transmit beamforming and RIS reflective beamforming were optimized jointly to maximize the minimum secrecy rate of the RIS-assisted networks under the assumption of discrete and continuous RIS phase shifts. In \cite{artifical-noise}, the authors demonstrated the benefit of utilizing artificial noise to enhance the secrecy rate of an RIS-assisted wireless communication system. In addition, an iterative algorithm was proposed to optimize the BS transmit beamforming and RIS reflective beamforming for maximizing the achievable secrecy rate.

In \cite{energy-efficiency}, the energy and spectral efficiency were maximized by alternatively optimizing the BS transmit beamforming and RIS phase shifts in a multi-user multi-input-single-output (MISO) downlink network. Specifically, a water-filling like approach was proposed to solve the power allocation problem and two approaches, i.e., the gradient search and sequential optimization, were developed to optimize the RIS phase shifts. In addition, numerical results showed that the utilization of RISs improved the energy efficiency when compared to the utilization of amplify-and-forward relays. In \cite{achievable-rate}, the RIS phase shifts were related to the amplitude of the reflection coefficient. By using this new phase shift model, the authors formulated a non-convex problem of maximizing the achievable rate of a MISO downlink network and employed an alternating optimization to obtain a low-complexity algorithm.

The performance of mmWave networks has been studied intensively in previous literatures. Stochastic geometry is a powerful tool for the performance analysis of the cellular networks. In the traditional cellular networks \cite{UK-sub-6}\cite{UK-mmWave}. In \cite{blockage}, the authors analyzed the LOS probability of BSs in the cellular networks where the blockages were distributed randomly. The blockages were modeled utilizing the rectangle boolean scheme where the length, orientation and width were uniformly distributed and the midpoints of the blockages were modeled as a PPP. By utilizing a distance-dependent function was to model the LOS probability of the BSs, the coverage probability and achievable rate of the mmWave cellular networks were derived in \cite{Bai}. In addition, the coverage probability under a special case was also obtained where a step function was utilized to approximate the LOS probability function.

To further increase the spectral efficiency and user fairness, non-orthogonal multiple access (NOMA) and RIS were integrated. In \cite{DCA}, the utilization of RISs in the NOMA-based networks was analyzed. The difference of convex algorithm and alternating optimization were used to minimize the system power consumption by optimizing the BS beamforming and RIS phase shifts. In \cite{Ding}, the authors analyzed the performance of the RIS-assisted NOMA networks under two phase shifting designs, i.e., the coherent and random phase shifting. The performance of the RIS-NOMA networks were compared with that of the relaying and RIS-OMA networks. In \cite{ris-noma}, a priority-oriented design was proposed to enhance the spectral efficiency of an RIS-assisted NOMA network. In addition, the closed-from expressions of the outage probability and ergodic rate were derived. Some works have studied the performance of the RIS-assisted networks. In \cite{Macro}, the authors considered a RIS-assisted network where the RISs were modeled utilizing the boolean model and obtained the probability that a RIS can provide an indirect path for a given pair of transmitter and receiver. In \cite{stochastic-large}, the achievable rate and energy efficiency were derived. In \cite{exploiting}, the authors derived the ratio of the blind spots to the entire area and the distribution of the path loss between the typical user and its serving BS. However, the interference is not considered.

\subsection{Contributions}
While the RISs can provided indirect line-of-sight (LOS) paths for the blocked links, it brings strong interference since it reflects the signals from its surrounding BSs. Therefore, it is important to investigate the effect of the additional interference caused by the RISs. In this paper, we consider a mmWave cellular network where the blockages are randomly distributed. Then we user the tools from stochastic geometry to analyze the network performance by taking the blockage effect into consideration.

The main contributions of the work are summarized as follows
\begin{enumerate}
\item We consider a RIS-assisted mmWave cellular network where the blockages are modeled by the boolean scheme and a fraction of the blockages are coated by RISs. In addition, both BSs and users are equipped with the uniform linear array (ULA) and two antenna schemes are adopted. In the first scheme, the beam directions of the BSs/users and RISs are aligned to the angle of departure (AoD) and angle of arrival (AoA) of the channels between the BSs/users and RISs. In the second scheme, the beam directions of the BSs and users are aligned to the AoD/AoA of the channels between the BSs and users. The beamforming gain under both antenna schemes are analyzed.
\item We adopted a stochastic geometry-based framework to analyze the performance of a RIS-assisted mmWave cellular network by modeling the distributions of the BSs and RISs as PPPs. The distributions of the distance between the typical user and its serving BS/RIS are obtained. In addition, the active probability of the BSs and RISs are analyzed by taking the effect of the blockage into consideration. Based on the results, the coverage probability of a RIS-assisted mmWave cellular network is obtained. The coverage probability under the special case where the blockage parameter is sufficiently low is also derived. Moreover, the expressions of the ASE and energy efficiency are presented.
\item The simulation demonstrates the impact of the key network parameters, such as the density of the BSs/RISs, the size of RISs on the coverage probability, ASE and energy efficiency. The simulation results for the traditional cellular networks without RISs are also provided for the sake of comparison. The results demonstrated that the deployment of the RISs can effectively improve the coverage probability of the mmWave cellular networks. In addition, there exists a tradeoff between the BSs and RIS.
\end{enumerate}

The reminder of this paper is organized as follows. In Section $\text{\uppercase\expandafter{\romannumeral2}}$, the system model is introduced. In Section $\text{\uppercase\expandafter{\romannumeral3}}$, the coverage probability of the RIS-assisted mmWave networks are derived. In Section $\text{\uppercase\expandafter{\romannumeral4}}$, the energy efficiency of the RIS-assisted mmWave networks are derived In Section $\text{\uppercase\expandafter{\romannumeral5}}$, numerical results are presented to show the impact of key system parameters on the performance metrics. Finally, conclusions are provided in Section $\text{\uppercase\expandafter{\romannumeral6}}$.

\section{System model}
\begin{figure}
  \centering
  \includegraphics[width=3.5in]{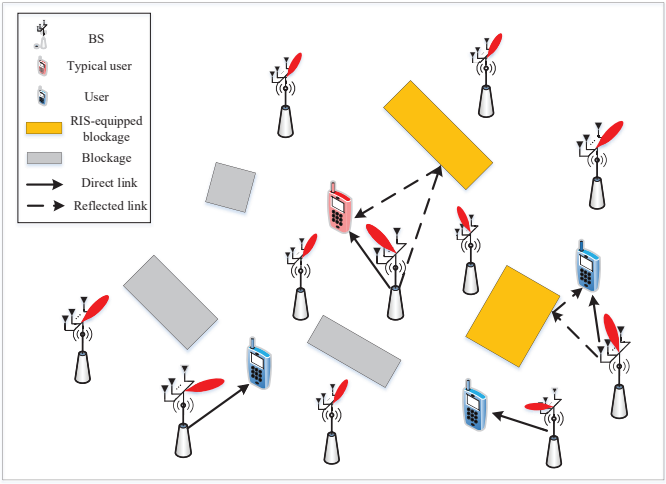}
  \caption{The layouts of a RIS-assisted mmWave network.}\label{RIS-framework}
\end{figure}

We consider a RIS-assisted cellular network, as shown in Fig. \ref{RIS-framework}. The locations of the BSs and users are assumed to follow an independent homogeneous Poisson Point Process (PPP) $\Phi_{BS}$ and $\Phi_u$ with density $\lambda_{BS}$ and $\lambda_u$, respectively. Each BS and user are equipped with a uniform linear array (ULA) and the number of antennas at the BSs and users are denoted by $N_{BS}$ and $N_u$, respectively. Let $P$ denote the transmit power of each BS. The blockage The available bandwidth is denoted by $W$ (in Hz). We assume that the blockages are distributed according to another independent homogeneous PPP $\Phi_{b}$ with density $\lambda_b$ while a faction of the blockage are coated with RISs. According to the thinning theorem \cite{thinning-theorem}, the RISs are also distributed as a PPP $\Phi_R$ with density $\lambda_R$. Without loss of generality, according to Slivnyak's theorem \cite{Bai}, we can study the performance of a typical user $u_0$ located at the origin $o\in\mathbb{R}$.

Due to that all BSs operate in the mmWave frequency, the communications link is sensitive to blockage effects. The link between $u_0$ and an arbitrary BS can be either line-of-sight (LOS) or non-line-of-sight (NLOS), which is dependent on the visibility of the BS to $u_0$. If there are no blockages between $u_0$ and the BS, the corresponding link is LOS. Otherwise, the link is considered to be NLOS. The LOS probability of an arbitrary link between $u_0$ and the BS at the distance $x$ is defined by a LOS probability function, i.e., $p(x)=e^{-\beta x}$, where the blockage parameter $\beta$ is dependent on the BS density and the blockages \cite{Bai}. The NLOS probability of the corresponding link is therefore $1-e^{-\beta x}$.

The desired signal received by $u_0$ is attenuated by the path loss and the severity of the attenuation can be measured by the path loss exponent. To distinguish the LOS/NLOS states of the link, different path loss laws are applied to the links under the LOS/NLOS link states. Therefore, the path loss gain is given by
\begin{equation}
L(x)=\mathbbm{1}(p(x))C_Lx^{-\alpha_L}+\mathbbm{1}(1-p(x))C_Nx^{-\alpha_N},
\end{equation}
where $C_{L}$ and $C_{N}$ are the path loss of the LOS/NLOS links at the reference distance of 1 meter, $\alpha_{L}$ and $\alpha_{L}$ represent the path loss exponents of the LOS and NLOS links, respectively.

Note that the signal is spatially sparse in the angle domain and the multipath component is mainly from reflections rather than scattering and refraction \cite{mmWave-channel-1}-\cite{mmWave-channel-3}. Therefore, we neglect the small-scale fading.

$u_0$ is associated with the BS providing the strongest signal power. Note that the serving BS may not always lie in the set including the closest BS due to the LOS/NLOS state of the BSs caused by the blockage. Hence, $u_0$ is associated with a BS based on the maximum biased received power criterion (Max-BRP).

We consider two types of channels, i.e., the direct channel and the reflected channel. The direct channel denotes the channel between the typical user $u_0$ and its serving BS. Each BS is equipped with the uniform linear array (ULA). We assume that the RIS is connected with the BSs and the phase shifters can be controlled at each time slot. The channel of the direct link can be expressed as
\begin{equation}
\mathbf{h}_{d,0}=\mathbf{a}_{r}(\phi_{d,0})^H\mathbf{a}_{t}(\theta_{d,0}).
\end{equation}

Note that the matrices $\alpha_{t}(\phi_{\text{BU},0})$ and $\alpha_{r}(\theta_{\text{BU},0})$ is
\begin{equation}
\begin{split}
&\mathbf{a}_{r}(\phi_{d,0})=[1,e^{j2\pi\varphi_{d,0}},\cdots,e^{j(N_{BS}-1)2\pi\varphi_{d,0}}],\\
&\mathbf{a}_{t}(\theta_{d,0})=[1,e^{j2\pi\vartheta_{d,0}},\cdots,e^{j(N_{u}-1)2\pi\vartheta_{d,0}}],
\end{split}
\end{equation}
where $\vartheta_{d,0}=(d/\omega)\theta_{d,0}$, $\varphi_{d,0}=(d/\omega)\phi_{d,0}$, $d$ denotes the distance between the antenna elements, $\omega$ the signal wavelength, $\phi_{d,0}\sim\text{U}(0,2\pi)$ and $\theta_{d,0}\sim\text{U}(0,2\pi)$ the angle of departure (AoD) at the BS and the angle of arrival (AoA) at the user, respectively. The matched filter optimal analog beamforming vector is applied at the users and BSs, which is shown as follows
\begin{equation}
\begin{cases}
\mathbf{w}_{r}(\phi_{0})=\sqrt{\frac{1}{N_u}}\mathbf{a}_{r}(\phi_{0})\\
\mathbf{w}_{t}(\theta_{0})=\sqrt{\frac{1}{N_{BS}}}\mathbf{a}_{t}(\theta_{0}).
\end{cases}
\end{equation}

We adopt two antenna schemes where the first scheme aims to optimize the reflected channel and the second scheme aims to optimize the direct channel. In the reminder of the paper, we analyze the network performance under antenna scheme 1 and the performance under antenna scheme 2 can be obtained following the similar steps. The antenna gain $M_{\text{DL},0}$ for the BS-user link is given by
\begin{equation}\label{antenna-gain-function}
\begin{split}
M_{d,0}=&\left|\mathbf{w}_{r}(\theta_{u,0})\mathbf{h}_{d,0}\mathbf{w}_{t}^H(\theta_{d,0})\right|^2\\
=&\left|\mathbf{w}_{r}(\theta_{u,0})\mathbf{a}_{r}(\phi_{d,0})^H\right|^2
\left|\mathbf{a}_{t}(\theta_{d,0})\mathbf{w}_{t}^H(\theta_{d,0})\right|^2\\
=&\frac{1}{N_{BS}N_u}\left|\sum_{i=0}^{N_{BS}}e^{j2\pi i(\vartheta_{d,0}-\vartheta_{g,0})}\right|^2\left|\sum_{i=0}^{N_u}e^{j2\pi i(\varphi_{d,0}-\varphi_{u,0})}\right|^2\\
=&\frac{1}{N_{BS}N_u}\frac{\sin^2\left(\pi N_{BS}(\vartheta_{d,0}-\vartheta_{g,0})\right)}{\sin^2(\pi(\vartheta_{d,0}-\vartheta_{g,0}))}\\
&\frac{\sin^2\left(\pi N_t(\varphi_{d,0}-\varphi_{u,0})\right)}{\sin^2(\pi(\varphi_{d,0}-\varphi_{u,0}))}.
\end{split}
\end{equation}

By defining the antenna gain for the BSs and users from the BS-user link as
\begin{equation}
m_{BS,\text{DL},0}=\frac{\sin^2\left(\pi N_{BS}(\vartheta_{d,0}-\vartheta_{g,0})\right)}{\sin^2(\pi(\vartheta_{d,0}-\vartheta_{g,0}))},
\end{equation}
\begin{equation}
m_{u,\text{DL},0}=\frac{\sin^2\left(\pi N_{u}(\varphi_{d,0}-\varphi_{u,0})\right)}{\sin^2(\pi(\varphi_{d,0}-\varphi_{u,0}))},
\end{equation}
where $m_{BS,d,0}$ and $m_{u,d,0}$ are Fej\'er kernels. Then the antenna gain $M_{d,0}$ for the BS-user link can be expressed as
\begin{equation}
M_{\text{DL},0}=m_{BS,\text{DL},0}m_{u,\text{DL},0}.
\end{equation}

Letting $\theta_{g,i_0}$ be the AoD of the channel between the interfering BS and RIS, the antenna gain $M_{d,i}$ for the interfering BS-user link can be derived as
\begin{equation}\label{antenna-gain-derivation}
\begin{split}
M_{\text{DL},i}=&\frac{1}{N_{BS}N_u}\underbrace{\frac{\sin^2\left(\pi N_{BS}(\vartheta_{d,i}-\vartheta_{g,i_0})\right)}{\sin^2(\pi(\vartheta_{d,i}-\vartheta_{g,i_0}))}}_{m_{BS,\text{dl}}}\\
&\underbrace{\frac{\sin^2\left(\pi N_t(\varphi_{d,i}-\varphi_{u,0})\right)}{\sin^2(\pi(\varphi_{d,i}-\varphi_{u,0}))}}_{m_{u,\text{dl}}}.
\end{split}
\end{equation}

Since $\theta_{d,0}$ and $\phi_{d,0}$ are assumed to follow the uniform distribution, we can derive the expectation of the antenna gain, i.e., $\mathbb{E}[M_{\text{DL},i}]$ and decompose it into the product of the antenna gains at the BS and user, i.e., $\mathbb{E}[M_{\text{DL},i}]=\bar{m}_{BS,\text{DL}}\bar{m}_{u,\text{DL}}$, where the average antenna gains at the BS and user are derived as
\begin{equation}
\begin{split}
\bar{m}_{BS,\text{DL}}&=\frac{1}{2\pi N_{BS}}\int_{0}^{2\pi}m_{BS}\text{d}\theta_{d,0}\\
&=\int_{0}^{2\pi}\frac{1-\cos\left(2\pi N_{BS}(\vartheta_{d,i}-\vartheta_{g,i_0})\right)}{1-\cos(2\pi(\vartheta_{d,i}-\vartheta_{g,i_0}))}\text{d}\theta_{d,0}.
\end{split}
\end{equation}
\begin{equation}
\bar{m}_{u,\text{DL}}=\frac{1}{2\pi N_u}\int_{0}^{2\pi}\frac{1-\cos\left(2\pi N_t(\varphi_{d,i}-\varphi_{u,0})\right)}{1-\cos(2\pi(\varphi_{d,i}-\varphi_{u,0}))}\text{d}\phi_{d,0}.
\end{equation}

The channel for the BS-RIS link can be expressed as
\begin{equation}
\mathbf{h}_{g,0}=\mathbf{a}_{r}(\phi_{g,0})^H\mathbf{a}_{t}(\theta_{g,0}).
\end{equation}

Note that the steering matrices can be expressed as
\begin{equation}
\begin{split}
&\mathbf{a}_{r}(\phi_{g,0})=[1,e^{j2\pi\varphi_{g,0}},\cdots,e^{j(N-1)2\pi\varphi_{g,0}}],\\
&\mathbf{a}_{t}(\theta_{g,0})=[1,e^{j2\pi\vartheta_{g,0}},\cdots,e^{j(N_{BS}-1)2\pi\vartheta_{g,0}}],
\end{split}
\end{equation}
where $\varphi_{g,0}=(d/\omega)\sin(\phi_{g,0})$, $\vartheta_{g,0}=(d/\omega)\sin(\theta_{g,0})$, $\phi_{g,0}\sim\text{U}(0,2\pi)$ and $\theta_{g,0}\sim\text{U}(0,2\pi)$ the angle of departure (AoD) at the BS and the angle of arrival (AoA) at the RIS for the BS-RIS link.

Similarly, the channel for the RIS-user link can be expressed as
\begin{equation}
\mathbf{h}_{u,0}=\mathbf{a}_{r}(\phi_{u,0})^H\mathbf{a}_{t}(\theta_{u,0}).
\end{equation}

Note that the steering matrices are expressed as
\begin{equation}
\begin{split}
&\mathbf{a}_{r}(\phi_{u,0})=[1,e^{j2\pi\varphi_{u,0}},\cdots,e^{j(N_u-1)2\pi\varphi_{u,0}}],\\
&\mathbf{a}_{t}(\theta_{u,0})=[1,e^{j2\pi\vartheta_{u,0}},\cdots,e^{j(N-1)2\pi\vartheta_{u,0}}],
\end{split}
\end{equation}
where $\varphi_{u,0}=(d/\omega)\sin(\phi_{u,0})$, $\vartheta_{u,0}=(d/\omega)\sin(\theta_{u,0})$, $\phi_{u,0}\sim\text{U}(0,2\pi)$ and $\theta_{u,0}\sim\text{U}(0,2\pi)$ the angle of departure (AoD) at the RIS and the angle of arrival (AoA) at the user for the RIS-user link.

The channel for the BS-user link via a RIS can be expressed as
\begin{equation}
\mathbf{h}_{R,0}=\mathbf{h}_{u,0}(\phi_{u,0})^H\mathbf{\Psi}_0\mathbf{h}_{g,0}(\theta_{u,0}).
\end{equation}

Note that $\mathbf{\Psi}_0$ is a diagonal matrix introduced by the RIS as
\begin{equation}
\mathbf{\Psi}_0=\text{diag}\{e^{j\psi_1},\cdots,e^{j\psi_{N,0}}\},
\end{equation}
where $\psi_{n,0}\in[0,2\pi)$ is the phase shift introduced by the $n$-th element of the RIS. By applying the matched filter optimal analog beamforming, the antenna gain $M_{\text{rl},0}$ for the serving BS-user link through a RIS is expressed as
\begin{equation}\label{antenna-gain-RIS}
\begin{split}
M_{\text{RL},0}=&\left|\mathbf{w}_{r}(\phi_{u,0})\mathbf{h}_{u,0}(\phi_{u,0})^H\mathbf{\Psi}_0\mathbf{h}_{g,0}(\theta_{u,0})\mathbf{w}_{t}(\theta_{g,0})\right|^2\\
=&\left|\mathbf{w}_{r}(\phi_{u,0})\mathbf{a}_{r}(\phi_{u,0})^H\right|^2
\left|\mathbf{a}_{t}(\theta_{u,0})\Psi_0\mathbf{a}_{r}(\phi_{g,0})\right|^2\\
&\left|\mathbf{a}_{t}(\theta_{g,0})\mathbf{w}_{t}(\theta_{g,0})^H\right|^2\\
=&N_uN_{BS}\left|\mathbf{a}_{t}(\theta_{u,0})\mathbf{\Psi}_0\mathbf{a}_{r}(\phi_{g,0})\right|^2.
\end{split}
\end{equation}

From (\ref{antenna-gain-RIS}), we can observe that $M_{g,i}$ depends on $\theta_{u,0}$ and $\phi_{g,0}$, but is independent of $\phi_{u,0}$ and $\theta_{g,0}$. Therefore, the phase shift of the RIS should be optimized to maximize $\left|\mathbf{a}_{t}(\theta_{u,0})\mathbf{\Psi}_0\mathbf{a}_{r}(\phi_{g,0})\right|$, which is rewritten as \cite{large-intelligent}
\begin{equation}\label{antenna-gain-maximization}
\begin{split}
&\left|\mathbf{a}_{t}(\theta_{u,0})\mathbf{\Psi}_0\mathbf{a}_{r}(\phi_{g,0})\right|^2\\
&=\mathbbm{E}\left[\left|\sum_{n=1}^{N}e^{j2\pi(n-1)(\vartheta_{u,0}-\varphi_{g,0})+j\psi_{n,0}}\right|^2\right].
\end{split}
\end{equation}

The optimal phase shift on the $n$-th element of the serving RIS is
\begin{equation}\label{optimal-phase-shift}
\psi_{n,0}=2\pi(n-1)(\vartheta_{u,0}-\varphi_{g,0}).
\end{equation}

By substituting (\ref{optimal-phase-shift}) into (\ref{antenna-gain-maximization}), the optimal antenna gain $M_{\text{rl},0}$ is $N_{BS}N_u N^2$. Letting $\theta_{g,i_0}$ be the AoA of the channel between $u_0$ and the interfering BS, the phase shift on the $n$-th element of the interfering RIS is
\begin{equation}
\psi_{n,i}=2\pi(n-1)(\vartheta_{u,i_0}-\varphi_{g,i_0}).
\end{equation}

The antenna gain $M_{\text{RL},i}$ for the interfering BS-user link via a RIS is expressed as
\begin{equation}
\begin{split}
&M_{\text{RL},i}=\left|\sum_{i=0}^{N_{BS}}\frac{e^{j2\pi i(\vartheta_{g,0}-\vartheta_{g,i_0})}}{N_{BS}}\right|^2\left|\sum_{n=0}^{N_u}\frac{e^{j2\pi i(\varphi_{u,0}-\varphi_{u,i_0})}}{N_u}\right|^2\\
&\left|\sum_{n=1}^{N}e^{j2\pi(n-1)(\vartheta_{u,i}-\varphi_{g,i})+j\psi_{n,0}}\right|^2\\
&=\underbrace{\frac{\sin^2\left(\pi N_{BS}\left(\vartheta_{g,i}-\vartheta_{g,i_0}\right)\right)}{N_{BS}\sin^2\left(\pi \left(\vartheta_{g,i}-\vartheta_{g,i_0}\right)\right)}}_{m_{BS,\text{RL}}}
\underbrace{\frac{\sin^2\left(\pi N_u\left(\varphi_{u,i}-\varphi_{u,i_0}\right)\right)}{N_u\sin^2\left(\pi \left(\varphi_{u,i}-\varphi_{u,i_0}\right)\right)}}_{m_{u,\text{RL}}}\\
&\underbrace{\frac{\sin^2\left(\pi N\left(\vartheta_{u,i}-\vartheta_{u,i_0}-\varphi_{g,i}+\varphi_{g,i_0}\right)\right)}{\sin^2\left(\pi \left(\vartheta_{u,i}-\vartheta_{u,i_0}-\varphi_{g,i}+\varphi_{g,i_0}\right)\right)}}_{m_{r,\text{RL}}}.
\end{split}
\end{equation}

Note that $\theta_{g,i}$, $\phi_{u,i}$, $\theta_{u,i}$, $\theta_{u,i_0}$, $\phi_{g,i}$ and $\phi_{g,i_0}$ are assumed to follow the uniform distribution. Hence, we derive the expectation of the antenna gain, i.e., $\mathbb{E}[M_{\text{RL},i}]$ and decompose it into the product of the antenna gains at the BS and user, i.e., $\mathbb{E}[M_{\text{RL},i}]=\bar{m}_{BS,\text{RL}}\bar{m}_{u,\text{RL}}\bar{m}_{r,\text{RL}}$. The average antenna gains at $u_0$ and the interfering BS for the reflected link can be obtained following the similar steps with that for the direct link. In addition, the average antenna gain at the RIS for the reflected link are derived as
\begin{equation}
\begin{split}
\bar{m}_{r,\text{RL}}=&\frac{1}{16\pi^4 N}\int_{0}^{2\pi}\int_{0}^{2\pi}\int_{0}^{2\pi}\int_{0}^{2\pi}\\
&m_{r,\text{RL}}\text{d}\theta_{u,i}\phi_{g,i}\theta_{u,i_0}\phi_{g,i_0}.
\end{split}
\end{equation}

\subsection{Performance Metric}
Assume that a BS is idle when no users are located in its Voronoi cell. Let $\tilde{\Phi}_{BS}$ denote the set of idle BSs. the received signal is given in (\ref{received-signal}), as shown on the top of the next page.
\begin{figure*}
\begin{equation}\label{received-signal}
\begin{split}
R=&\left(\sqrt{PL_{d,0}}\mathbf{h}_{d,0}^{H}+\sqrt{PL_{g,0}L_{u,0}}\mathbf{h}_{g,0}^{H}\right)\mathbf{w}_0s_0
+\sum_{i\in\tilde{\Phi}_{BS}}\sqrt{PL_{d,i}}\mathbf{h}_{d,i}^{H}\mathbf{w}_is_i
+\sum_{i\in\tilde{\Phi}_R}\sqrt{PL_{u,0}}\mathbf{h}_{g,0}^{H}\Theta_i\sum_{m_i\in\bar{\Phi}_{BS}}L_{g,m_i}\mathbf{w}_{m_i}s_{m_i}\\
&+\sum_{i\in\Phi_R/\tilde{\Phi}_R}\sqrt{PL_{u,0}}\mathbf{h}_{g,0}^{H}\Theta_{i,\text{idle}}\sum_{j_i\in\bar{\Phi}_{BS}}L_{g,j_i}\mathbf{w}_{j_i}s_{j_i}.
\end{split}
\end{equation}
\end{figure*}
where $x_0$ is the transmitted signal from the serving BS to $u_0$, $x_{i}$ the transmitted signals from the interfering BSs to their associated users, $L_{d,0}$ the path loss gain for the BS-user link, $L_{u,0}$ the path loss gain of the RIS-user link, $L_{g,0}$ the path loss gain of the BS-RIS link and $n_{0}\sim\mathcal{CN}(0,\sigma^2)$ the complex additive white Gaussian noise (AWGN). Then the signal-to-interference-plus-noise (SINR) can be written as
\begin{equation}
\text{SINR}=\frac{\left|\left(\sqrt{PL_{d,0}M_{\text{DL},0}}+\sqrt{PL_{g,0}L_{u,0}M_{\text{RL},0}}\right)\right|^2}
{I_{BS}+I_{R}+I_{R,\text{idle}}+\sigma^2},
\end{equation}
where $I_{BS}=\sum_{i\in\tilde{\Phi}_{BS}}PL_{d,i}\left|\mathbf{h}_{d,i}^{H}\mathbf{w}_i\right|^2$,
$I_{R}=\sum_{i\in\tilde{\Phi}_R}PL_{u,0}\left|\mathbf{h}_{g,0}^{H}\Theta_i\sum_{m_i\in\bar{\Phi}_{BS}}L_{g,m_i}\mathbf{w}_{m_i}\right|^2$,
$I_{R,\text{idle}}=\sum_{i\in\Phi_R/\tilde{\Phi}_R}PL_{u,0}\left|\mathbf{h}_{g,0}^{H}\Theta_{i,\text{idle}}\sum_{j_i\in\bar{\Phi}_{BS}}L_{g,j_i}\mathbf{w}_{j_i}\right|^2$.

In this paper, we adopt the coverage probability \cite{Bai} as the performance metric. The transmission process is considered to be successful if the SINR surpasses the threshold $T$. By the law of total probability, the coverage probability is given by
\begin{equation}\label{average load}
\begin{split}
\mathcal{P}(T)=&\mathbb{P}\left(\text{SINR}\geq T\right),
\end{split}
\end{equation}

For ease of reference, a list of notations and parameters used in this paper is presented in Table I.
\newcommand{\tabincell}[2]{\begin{tabular}{@{}#1@{}}#2\end{tabular}}  

\begin{table*}[htbp]
\centering
\caption{\label{tab:test}Notations and Parameters}
\begin{tabular}{c|cc}
\hline
\hline
$\textbf{Notation}$&$\textbf{Parameter}$\\
\hline
$P$ & \tabincell{c}{The transmit power of the BSs} \\
\hline
$\Phi_{BS};\Phi_R;\Phi_u$ & The set of the BSs/RISs/users\\
\hline
$\lambda_{BS};\lambda_R;\Phi_u$ & The density of the BSs/RISs/users\\
\hline
$l_{\rho,z}$ & \tabincell{c}{The path loss between the user and the LOS/NLOS BS}\\
\hline
$\alpha_L;\alpha_N$ & \tabincell{c}{The path loss exponent of the channel with LOS/NLOS BS} \\
\hline
$\beta$ & \tabincell{c}{The blockage parameter}\\
\hline
$C_L;C_N$ & \tabincell{c}{The path loss of the LOS/NLOS channel at the reference distance of 1 meter}\\
\hline
$\theta_{d,0};\phi_{d,0};\theta_{d,i};\phi_{d,i}$ & \tabincell{c}{The AoD/AoA of the link between the serving/interfering BS and $u_0$}\\
\hline
$\theta_{u,0};\phi_{u,0};\theta_{u,i};\phi_{u,i}$ & \tabincell{c}{The AoD/AoA of the link between the serving/interfering RIS and $u_0$}\\
\hline

\end{tabular}
\end{table*}

\section{Analysis of coverage probability}
In this section, we first provide the characterization of the path loss, i.e., the probability density function (PDF) and the complementary cumulative distribution function (CCDF) and of the distance between $u_0$ and its serving BS/RIS, then derive the probability of $u_0$ being associated with a LOS/NLOS link.

The PDF of the distance between $u_0$ and its nearest BS is
\begin{equation}
\hat{f}_{d}(x)=2\pi\lambda_{BS}re^{-2\pi\lambda_{BS}\int_{0}^{x}r\text{d}x}.
\end{equation}

Accordingly, the PDF of the distance between $u_0$ and its nearest LOS BS is
\begin{equation}
\hat{f}_{d,L}(x)=2\pi\lambda_{BS}p(r)re^{-2\pi\lambda_{BS}\int_{0}^{x}p(r)r\text{d}x}.
\end{equation}

\begin{Lemma}\label{lemma-1}
The probability that $u_0$ is associated with a LOS BS is
\begin{equation}\label{association-los}
A_{d,L}(x)=\int_{0}^{\infty}\hat{f}_{d,L}(x)e^{-(2\pi\lambda_{BS}\int_{0}^{\chi_{L}(x)}(1-p(r))r\text{d}r)}\text{d}x,
\end{equation}
where $\chi_{L}(x)=(C_N/C_L)^{1/\alpha_N}x^{\alpha_L/\alpha_N}$.

\emph{Proof:} See Appendix A.
\end{Lemma}

By utilizing the above results, the PDF of the distance between $u_0$ and its serving BS can be obtained in the following lemma.
\begin{Lemma}
Given that $u_0$ is associated with a LOS BS, the PDF of the distance between $u_0$ and its serving LOS BS can be expressed as
\begin{equation}
f_{d,L}(x)=\frac{\hat{f}_{d,L}(x)}{A_{d,L}(x)}e^{-(2\pi\lambda_{BS}\int_{0}^{\chi_{L}(x)}(1-p(r))r\text{d}r)}.
\end{equation}
\end{Lemma}

Accordingly, the PDF of the distance between $u_0$ and its nearest NLOS BS is
\begin{equation}
\hat{f}_{d,N}(x)=2\pi\lambda_{BS}(1-p(r))re^{-2\pi\lambda_{BS}\int_{0}^{\chi_{L}(x)}(1-p(r))r\text{d}x}.
\end{equation}

The probability that $u_0$ is associated with a NLOS BS is
\begin{equation}\label{association-nlos}
A_{d,N}(x)=\int_{0}^{\infty}\hat{f}_{d,N}(x)e^{-(2\pi\lambda_{BS}\int_{0}^{\chi_{N}(x)}p(r)r\text{d}r)}\text{d}x.
\end{equation}

From (\ref{association-los}) and (\ref{association-nlos}), we can observe that the association probability is dependent on the BS density and blockage parameter.

Given that $u_0$ is associated with a NLOS BS, the PDF of the distance between $u_0$ and its serving NLOS BS can be expressed as
\begin{equation}
A_{d,N}(x)=\frac{\hat{f}_{d,N}(x)}{A_{\text{BU},N}(x)}e^{-(2\pi\lambda_{BS}\int_{0}^{\chi_{L}(x)}p(r)r\text{d}r)}.
\end{equation}

\begin{figure}
  \centering
  \includegraphics[width=2.5in]{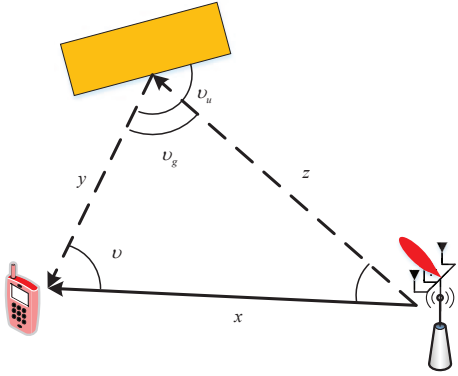}
  \caption{The geometric relationship between the direct and reflected links.}\label{RIS-link}
\end{figure}
As depicted in Fig. \ref{RIS-link}, different from the BS case, $u_0$ is associated with a assisting RIS when $u_0$ faces the RIS-equipped side of the blockage and $u_0$ and its serving BS are on the same side of the RIS. Then we derive the PDF of the distance between $u_0$ and its nearest LOS RIS as
\begin{equation}
\hat{f}_{u,L}(x,y,\upsilon)=\pi\lambda_{R}\mathcal{C}(x,y,\upsilon)p(r)re^{-\pi\lambda_{R}\mathcal{C}(x,y,\upsilon)\int_{0}^{y}p(r)r\text{d}x}.
\end{equation}

\begin{Lemma}\label{lemma-2}
The probability that $u_0$ is associated with a LOS RIS can be derived as
\begin{equation}\label{association-ris}
\begin{split}
&A_{u,L}(x,y,\upsilon)\\
&=\int_{0}^{\infty}\hat{f}_{u,L}(y)e^{-(\pi\lambda_{R}\mathcal{C}(x,y,\upsilon)\int_{0}^{\chi_{L}(y)}(1-p(r))r\text{d}r)}\text{d}x,
\end{split}
\end{equation}
where $\chi_{L}(y)=(C_N/C_L)^{1/\alpha_N}y^{\alpha_L/\alpha_N}$.

\emph{Proof:} See Appendix B.
\end{Lemma}

From (\ref{association-ris}), we can observe that the association probability is not only dependent on the BS/RIS density and blockage parameter but also the angle between the BS-user and RIS-user link.

Given that $u_0$ is associated with a LOS RIS, the PDF of the distance between $u_0$ and its serving LOS RIS can be expressed as
\begin{equation}
A_{u,L}(x,y,\upsilon)=\frac{\hat{f}_{u,L}(x,y,\upsilon)}{A_{u,L}(x,y,\upsilon)}e^{-(2\pi\lambda_R\mathcal{C}(x,y,\upsilon)\int_{0}^{\chi_{L}(y)}p(r)r\text{d}r)}.
\end{equation}

Note that the PDF of the distance between $u_0$ and its serving LOS RIS can be obtained following the similar steps. When the link between $u_0$ and its serving RIS and the link between the RIS and the BS are both LOS, $u_0$ is considered to be associated with a LOS BS through a RIS. Otherwise, $u_0$ is associated with a NLOS BS through the RIS. Based on the above definitions, we then analyze the association probability in the following lemma.
\begin{Lemma}
The probability that $u_0$ is associated with a LOS BS through a RIS is
\begin{equation}
\begin{split}
A_{g,L}(x)=&A_{u,L}(x)\int_{0}^{2\pi}\int_{0}^{\infty}\int_{0}^{\infty}\frac{1}{2\pi}f_{d}(x)f_{u,L}(y)\\
&p(\sqrt{x^2+y^2-2\cos\phi})\text{d}x\text{d}y\text{d}\phi.
\end{split}
\end{equation}

Accordingly, the probability that $u_0$ is associated with a NLOS BS through the RIS is
\begin{equation}
\begin{split}
A_{g,N}(x)=&A_{u,L}(x)\int_{0}^{2\pi}\int_{0}^{\infty}\int_{0}^{\infty}\frac{1}{2\pi}f_{d}(x)f_{u,L}(y)\\
&(1-p(\sqrt{x^2+y^2-2\cos\phi}))\text{d}x\text{d}y\text{d}\phi+A_{u,N}(x).
\end{split}
\end{equation}
\end{Lemma}

When the density of users are comparable or not much larger than that of the BSs or the RISs, some BSs or RISs will be idle. Specifically, a BS or RIS is idle if there are no users located within its Voronoi cell. Let the active probability of the BSs be $p_{a,BS}$. According to the thinning theorem, the distribution of the active BSs is a PPP with density $p_{a,BS}\lambda_{BS}$. We then derive the active probability of the BSs as
\begin{equation}
\begin{split}
&p_{a,BS}=1-\int_{0}^{\infty}e^{-\lambda_ux}f_{\mathcal{S}_{BS}}(x)\text{d}x\approx 1-\left(1+\frac{\lambda_u}{\lambda_{BS}}\right)^{-3.5},
\end{split}
\end{equation}
where $f_{\mathcal{S}_{BS}}(x)$ is the PDF of the area $\mathcal{S}_{BS}$. The area of a Voronoi cell in the Poisson random tessellation is
\begin{equation}
f_{\mathcal{S}_{BS}}(x)=\frac{3.5^{3.5}}{\Gamma(3.5)}\left(\lambda_{BS}\right)^{3.5}x^{2.5}e^{-3.5\lambda_{BS}x}.
\end{equation}

Note that the active probability of the BSs $p_{a,BS}$ approaches to 0 when $\lambda_u>>\lambda_{BS}$. Similarly, the empty cell probability of the RISs can be derived as
\begin{equation}
\begin{split}
&p_{a,R}=1-\int_{0}^{\infty}e^{-\lambda_ux}f_{\mathcal{S}_{R}}(x)\text{d}x\approx 1-\int_{0}^{2\pi}\int_{0}^{\infty}\int_{0}^{\infty}\frac{1}{2\pi}\\
&\left(1+\frac{2\lambda_u}{\mathcal{C}(x,y,\upsilon)\lambda_{R}}\right)^{-3.5}
f_{d}(x)f_{u}(y)\text{d}x\text{d}y\text{d}\upsilon.
\end{split}
\end{equation}

Next, we continue to analyze the coverage probability. According to the LOS/NLOS link state, the interference experienced by $u_0$ are from four sets, i.e., the set of LOS BSs $\Phi_{BS,L}$, the set of NLOS BSs $\Phi_{BS,N}$, the set of LOS RISs $\Phi_{R,L}$ and the set of NLOS RISs $\Phi_{R,N}$. By carefully handling the interference from the four sets, the coverage probability can be derived utilizing the law of total probability in the following theorem.
\begin{Theorem}\label{theorem-1}
The coverage probability of a RIS-assisted mmWave network under antenna scheme 1 is given by
\begin{equation}\label{STP-general-case}
\begin{split}
&\mathcal{P}_1(T)=\frac{1}{2\pi}\sum_{\rho\in\{L,N\}}\sum_{\xi\in\{L,N\}}\int_{0}^{2\pi}\int_{0}^{\infty}\int_{0}^{\infty}\sum_{w=1}^{W}
(-1)^{w+1}\binom{W}{w}\\
&\mathcal{L}_{I_{BS,L}}(s_{\rho,\xi})\mathcal{L}_{I_{BS,N}}(s_{\rho,\xi})
\mathcal{L}_{I_{R,L}}(s_{\rho,\xi})\mathcal{L}_{I_{R,N}}(s_{\rho,\xi})\mathcal{L}_{I_{R,L,\text{idle}}}(s_{\rho,\xi})\\
&\mathcal{L}_{I_{R,N,\text{idle}}}(s_{\rho,\xi})e^{-s_{\rho,\xi}\sigma^2}f_{d,\rho}(x)f_{u,\xi}(y)\text{d}x\text{d}y\text{d}\upsilon,
\end{split}
\end{equation}
where
\begin{equation}
\begin{split}
&\mathcal{L}_{I_{BS,L}}(s_{\rho,\xi})=\exp\left(-2\pi\lambda_{BS}p_{a,BS}\int_{\chi_{L,\rho}(x)}^{\infty}e^{-\beta r}\right.\\
&\left.\left(1-e^{-s_{\rho}P\bar{m}_{BS,\text{DL}}\bar{m}_{u,\text{DL}}C_Lr^{-\alpha_L}}\right)r\text{d}r\right),
\end{split}
\end{equation}
\begin{equation}
\begin{split}
&\mathcal{L}_{I_{BS,N}}(s_{\rho,\xi})=\exp\left(-2\pi\lambda_{BS}p_{a,BS}\int_{\chi_{N,\rho}(x)}^{\infty}(1-e^{-\beta r})\right.\\
&\left.\left(1-e^{-s_{\rho}P\bar{m}_{BS,\text{DL}}\bar{m}_{u,\text{DL}}C_Nr^{-\alpha_N}}\right)r\text{d}r\right),
\end{split}
\end{equation}
\begin{equation}
\begin{split}
&\mathcal{L}_{I_{R,L}}(s_{\rho,\xi})=\exp\left(-\pi\lambda_Rp_{a,R}\int_{\chi_{L,\xi}(y)}^{\infty}e^{-\beta r}\right.\\
&\left.\left(1-e^{-s_{\rho}P_R\bar{m}_{r,\text{RL}}\bar{m}_{BS,\text{RL}}\bar{m}_{u,\text{RL}}C_Lr^{-\alpha_L}}\right)r\text{d}r\right),
\end{split}
\end{equation}
\begin{equation}
\begin{split}
&\mathcal{L}_{I_{R,N}}(s_{\rho,\xi})=\exp\left(-\pi\lambda_Rp_{a,R}\int_{\chi_{N,\xi}(y)}^{\infty}(1-e^{-\beta r})\right.\\
&\left.\left(1-e^{-s_{\rho}P_R\bar{m}_{r,\text{RL}}\bar{m}_{BS,\text{RL}}\bar{m}_{u,\text{RL}}C_Nr^{-\alpha_N}}\right)r\text{d}r\right),
\end{split}
\end{equation}
\begin{equation}
\begin{split}
&\mathcal{L}_{I_{R,L,\text{idle}}}(s_{\rho,\xi})=\exp\left(-\pi\lambda_Rp_{a,R}\int_{\chi_{L,\xi}(y)}^{\infty}e^{-\beta r}\right.\\
&\left.\left(1-e^{-s_{\rho}P_R\bar{m}_{r,\text{RL},\text{idle}}\bar{m}_{BS,\text{RL}}\bar{m}_{u,\text{RL}}C_Lr^{-\alpha_L}}\right)r\text{d}r\right),
\end{split}
\end{equation}
\begin{equation}
\begin{split}
&\mathcal{L}_{I_{R,N,\text{idle}}}(s_{\rho,\xi})=\exp\left(-\pi\lambda_Rp_{a,R}\int_{\chi_{N,\xi}(y)}^{\infty}(1-e^{-\beta r})\right.\\
&\left.\left(1-e^{-s_{\rho}P_R\bar{m}_{r,\text{RL},\text{idle}}\bar{m}_{BS,\text{RL}}\bar{m}_{u,\text{RL}}C_Nr^{-\alpha_N}}\right)r\text{d}r\right),
\end{split}
\end{equation}
where $s_{\rho}=\frac{w\epsilon T}{\left|\left(\sqrt{PL_{d,0}M_{\text{DL},0}}+\sqrt{PL_{g,0}L_{u,0}M_{\text{RL},0}}\right)\right|^2}$, $\epsilon=W(W!)^{\frac{1}{W}}$, $P_R=\pi\lambda_{BS}\int_{0}^{\infty}\left(C_Lz^{-\alpha_L}e^{-\beta z}+C_Nz^{-\alpha_N}(1-e^{-\beta z})\right)\text{d}z$. Note that $\chi_{L,\rho}(x)$ denotes $x$ when $\rho=L$ and $\chi_L(x)$ when $\rho=N$. $\chi_{N,\rho}(x)$ denotes $\chi_N(x)$ when $\rho=L$ and $x$ when $\rho=N$. $\chi_{L,\rho}(y)$ denotes $y$ when $\rho=L$ and $\chi_L(y)$ when $\rho=N$. $\chi_{N,\rho}(y)$ denotes $\chi_N(y)$ when $\rho=L$ and $y$ when $\rho=N$. Note that the coverage probability under antenna scheme 2 $\mathcal{P}_2$ can be derived following the similar method to that under antenna scheme 1. The coverage probability for the direct link $\mathcal{P}_d$ can be obtained by replacing $s_{\rho}$ with $\frac{w\epsilon T}{PL_{d,0}M_{\text{DL},0}}$.

\emph{Proof:} See Appendix C.
\end{Theorem}

Theorem 1 shows that the coverage probability is dependent on the physical layer parameters, i.e., BS densities of BSs and RISs, transmit powers and the path loss exponents. In addition, the coverage probability is related to the size of the RISs. When the density of users is far larger than that of BSs or RISs, the active probability of the BSs/RISs become 1.

From Theorem 1, we can also observe that the expression of coverage probability is not of a closed-form and it is difficult to obtain the analytical results. To facilitate the computation process, the Gauss-Chebyshev quadrature (GCQ) formula \cite{GCQ} is utilized to compute the coverage probability. Under this situation, the coverage probability can be expressed as follows
\begin{equation}\label{STP-GQ}
\begin{split}
&\mathcal{P}_1(T)=\frac{1}{2\pi}\sum_{w=1}^{W}(-1)^{w+1}\binom{W}{w}\sum_{q_1=1}^{Q_1}\sum_{q_2=1}^{Q_2}\sum_{q_3=1}^{Q_3}w_{q_1}w_{q_2}w_{q_3}\\
&\mathcal{L}_{I_{BS,L}}(s_{\rho,\xi})\mathcal{L}_{I_{BS,N}}(s_{\rho,\xi})
\mathcal{L}_{I_{R,L}}(s_{\rho,\xi})\mathcal{L}_{I_{R,N}}(s_{\rho,\xi})\mathcal{L}_{I_{R,L,\text{idle}}}(s_{\rho,\xi})\\
&\mathcal{L}_{I_{R,N,\text{idle}}}(s_{\rho,\xi})e^{-s_{\rho,\xi}\sigma^2}f_{d,\rho}(x)f_{u,\xi}(y)\text{d}x\text{d}y\text{d}\upsilon,
\end{split}
\end{equation}
where the abscissas and weight can be expressed as
\begin{equation}
x_{q_1}=\tan\left(\frac{\pi}{4}\cos\left(\frac{2q_1-1}{2Q_1}\pi\right)+\frac{\pi}{4}\right),
\end{equation}
\begin{equation}
y_{q_2}=\tan\left(\frac{\pi}{4}\cos\left(\frac{2q_2-1}{2Q_2}\pi\right)+\frac{\pi}{4}\right),
\end{equation}
\begin{equation}
\phi_{R,q_3}=\frac{1}{2}\left(\cos\left(\frac{2q_3-1}{Q_3}+1\right)+1\right),
\end{equation}
\begin{equation}
w_{q_1}=\frac{\pi^2\sin\left(\frac{2q_1-1}{2Q_1}\pi\right)}{4Q_1\cos^2\left[\frac{\pi}{4}\cos\left(\frac{2q_1-1}{2Q_1}\pi\right)+\frac{\pi}{4}\right]},
\end{equation}
\begin{equation}
w_{q_2}=\frac{\pi^2\sin\left(\frac{2q_2-1}{2Q_2}\pi\right)}{4Q_1\cos^2\left[\frac{\pi}{4}\cos\left(\frac{2q_2-1}{2Q_2}\pi\right)+\frac{\pi}{4}\right]}.
\end{equation}

\begin{Remark}
The characterization of (\ref{STP-general-case}) involves multi-integrals and achieving the numerical results needs considerable time. In order to reduce the computation complexity, we obtain a finite sum approximation by using the GCQ formula. The convergence rate can also be fast since only limited number of terms may be required for the computation to converge to an accurate result. However, the value of $\tau$ needs to be chosen appropriately in order to strike a balance between the complexity and accuracy.
\end{Remark}

Next, in order to make it more straightforward to observe the impact of network parameters on the coverage probability, we provide a closed-form expression of coverage probability under a special case where the blockage parameter is sufficiently small.
\begin{Corollary}\label{corollary-1}
When the blockage parameter is sufficiently small, the coverage probability of a RIS-assisted mmWave cellular network is given by
\begin{equation}\label{STP-general-case}
\begin{split}
&\mathcal{P}_{1,\infty}(T)=\frac{1}{2\pi}\sum_{\rho\in\{L,N\}}\int_{0}^{2\pi}\int_{0}^{\infty}\int_{0}^{\infty}\sum_{w=1}^{W}
(-1)^{w+1}\binom{W}{w}\\
&\mathcal{L}_{I_{BS}}(s_{\rho})
\mathcal{L}_{I_{R}}(s_{\rho})\mathcal{L}_{I_{R,\text{idle}}}(s_{\rho,\xi})
e^{-s_{\rho,\xi}\sigma^2}f_{d,\rho}(x)f_{u,\xi}(y)\text{d}x\text{d}y\text{d}\upsilon,
\end{split}
\end{equation}
where
\begin{equation}
\begin{split}
&\mathcal{L}_{I_{BS,L}}(s_{\rho})=\exp\left(-2\pi\lambda_{BS}p_{a,BS}\right.\\
&\left.\int_{x}^{\infty}e^{-\beta r}\left(1-e^{-s_{\rho}P\bar{m}_{BS}\bar{m}_uC_Lr^{-\alpha_L}}\right)r\text{d}r\right),
\end{split}
\end{equation}
\begin{equation}
\begin{split}
&\mathcal{L}_{I_{R,L}}(s_{\rho})=\exp\left(-\pi\lambda_Rp_{a,R}\right.\\
&\left.\int_{y}^{\infty}e^{-\beta r}\left(1-e^{-s_{\rho}P_R\bar{m}_{R}\bar{m}_g\bar{m}_uC_Lr^{-\alpha_L}}\right)r\text{d}r\right),
\end{split}
\end{equation}
\begin{equation}
\begin{split}
&\mathcal{L}_{I_{R,L,\text{idle}}}(s_{\rho})=\exp\left(-\pi\lambda_Rp_{a,R}\right.\\
&\left.\int_{y}^{\infty}e^{-\beta r}\left(1-e^{-s_{\rho}P_R\bar{m}_{R,\text{idle}}\bar{m}_g\bar{m}_uC_Lr^{-\alpha_L}}\right)r\text{d}r\right).
\end{split}
\end{equation}
\end{Corollary}

\section{Analysis of Energy Efficiency}
In this section, we first derive the area spectral efficiency. The ASE is defined as the product of the BS density and the BS throughput. We assume that antenna scheme 1 is adopted and the ASE under antenna scheme 2 can be obtained following the similar steps. According to the BS and RIS density, the analysis of the ASE can be divided into two cases. In the first case where $\lambda_{BS}p_{a,BS}>\lambda_{R}p_{a,R}$, the ASE is given by
\begin{equation}\label{area-spectral-efficiency}
\begin{split}
\zeta=&\lambda_{R} p_{a,R}\mathcal{P}_1(T)\log(1+T)\\
&+(\lambda_{BS} p_{a,BS}-\lambda_{R}p_{a,R})\mathcal{P}_d(T)\log(1+T).
\end{split}
\end{equation}

In the second case where $\lambda_{BS}p_{a,BS}<\lambda_{R}p_{a,R}$, the ASE is given by
\begin{equation}\label{area-spectral-efficiency}
\begin{split}
\zeta=\lambda_{BS} p_{a,BS}\mathcal{P}_1(T)\log(1+T).
\end{split}
\end{equation}

In practice, a main proportion of the energy for cellular networks is consumed by the RISs and BSs. The total BS power consumption consists of two components: the static power consumption and the transmit power. Henceforth, a linear approximation model has been widely employed to describe the power consumption, i.e., $P_{tot}=P_0+\Delta P$, where $1/\Delta$ denotes power amplifier efficiency, $P$ the transmit power and $P_0$ the static circuit power consumption. In addition, a BS is muted when there are no users to serve in its coverage and the corresponding power consumption is 0. In other words, the power consumption of active BSs with density $p_{a,BS}\lambda_{BS}$ is considered here. Therefore, the average power consumption for BSs is given by
\begin{equation}\label{power}
P_{avg,BS}=\lambda_{BS} p_{a,BS}({P_{0}}+\Delta P).
\end{equation}

The energy efficiency is then defined as the ratio of the ASE and the corresponding average power consumption:
\begin{equation}
\begin{split}
\text{EE}=\frac{\zeta}{P_{avg,BS}+NP_e},
\end{split}
\end{equation}
where $P_{avg,BS}$ and $\zeta$ are given by (\ref{power}) and (\ref{area-spectral-efficiency}), respectively.

\section{Simulation Results}
In this section, we consider a RIS-assisted mmWave cellular network. We present the the impact of the key networks parameters on the coverage probability, ASE and energy efficiency. The tradeoff between the BS and RIS density is also investigated. Unless otherwise stated, the parameters are set as listed in the following table.

\begin{table}[htbp]
\centering
\caption{\label{tab:test}System Parameters}
\begin{tabular}{|c|c|c|}
\hline
$\textbf{Parameters}$&$\textbf{Values}$\\
\hline
BS transmit power & \tabincell{c}{$P_{BS}=33$dBm}\\
\hline
BS static power consumption & $P_0=10$W\\
\hline
Efficiency of power amplifier & 1/$\Delta$=1/6\\
\hline
Number of antenna in BS & $N_{BS}=8$\\
\hline
Number of antenna in BS & $N_u=4$\\
\hline
Number of elements in RIS & $M=128$\\
\hline
\tabincell{c}{Power consumption\\ of each elements in RIS} & $P_e$=7dBm\\
\hline
Path loss exponent & $\alpha_{L}=2$, $\alpha_{N}=4$ \\
\hline
BS density & $\lambda_{BS}=10/(500^2\pi)$\\
\hline
RIS density & $\lambda_R=10/(500^2\pi)$\\
\hline
User density & $\lambda_u=100/(500^2\pi)$\\
\hline
Blockage parameter & \tabincell{c}{$\beta_1=0.01$}\\
\hline
Carrier frequency & 28GHz\\
\hline
Path loss intercept &$C_L=C_N=(F_c/4\pi)^2$\\
\hline
Noise figure & 10dB\\
\hline
Noise power & \tabincell{c}{-174dBm/Hz+10$\log_{10}$W\\+10dB}\\
\hline
\end{tabular}
\end{table}

\begin{figure}
  \centering
  \includegraphics[width=3.5in]{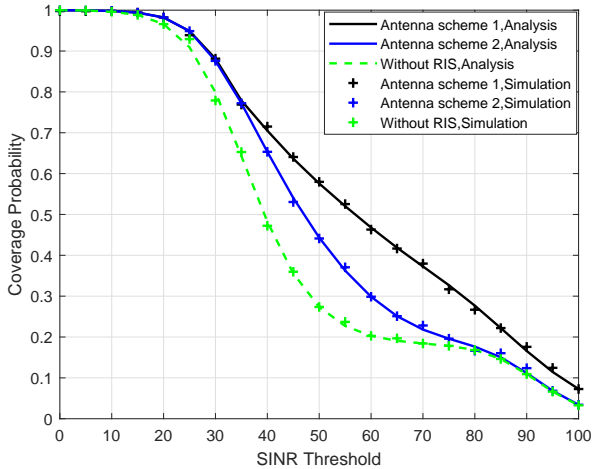}
  \caption{Coverage probability as functions of SINR threshold.}\label{cp-SINR-threshold}
\end{figure}

Fig. \ref{cp-SINR-threshold} illustrates the coverage probability as functions of the SINR threshold. It is observed that the coverage probability decreases with the SINR threshold. The coverage probability in the RIS-assisted networks is higher than that in the networks without RISs and the RIS-assisted networks under the antenna scheme 1 outperforms that under the antenna scheme 2. This can be explained as follows. The serving RIS is nearer than the serving BS to $u_0$. Hence, higher performance gain can be obtained when the main lobe is directed towards the RIS. In addition, it is shown that the theoretical results match the simulation results well, which verifies the correctness of the theoretical analysis.

\begin{figure}
  \centering
  \includegraphics[width=3.5in]{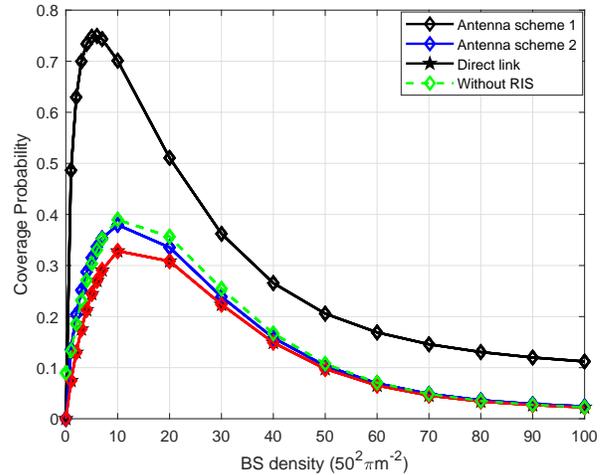}
  \caption{Coverage probability as functions of BS density.}\label{cp-BS-density}
\end{figure}

Fig. \ref{cp-BS-density} plots the coverage probability against the BS density. It can be observed that the coverage probabilities under three conditions initially increase with the increasing BS density. When the BS density is further increased, the coverage probabilities under three conditions decrease gradually. The results show that the network densification does not always improve the network performance. When the BS density is relatively small, the mmWave cellular network will benefit from the network densification. However, further improving the BS density will harm the network performance, especially when the BS density is sufficiently large. This is because $u_0$ is more likely to be interfered by the LOS BSs when the BS density is increased.

It can also be observed that the coverage probability is improved by introducing RISs into the networks and the coverage probability under antenna scheme 1 $\mathcal{P}_1$ is larger than that under antenna scheme 2 $\mathcal{P}_2$. Specifically, the gap between $\mathcal{P}_1$ and the coverage probability of the tradition cellular networks $\mathcal{P}_t$ becomes smaller when increasing the BS density and $\mathcal{P}_2$ is lower than $\mathcal{P}_t$. In addition, the coverage probability for the direct link, i.e., $\mathcal{P}_d$ is lowest. This can be explained as follows. When the RISs are deployed, the received signal strength is enhanced due to that a RIS may be utilized to provide a supplementary link to $u_0$. Moreover, as the RIS density is larger than the BS density, the distance between $u_0$ and the assisting RIS is shorter than that between $u_0$ and the serving BS. Then the performance gain is larger when the main lobe is directed into the channel between $u_0$ and the assisting RIS.

\begin{figure}
  \centering
  \includegraphics[width=3.5in]{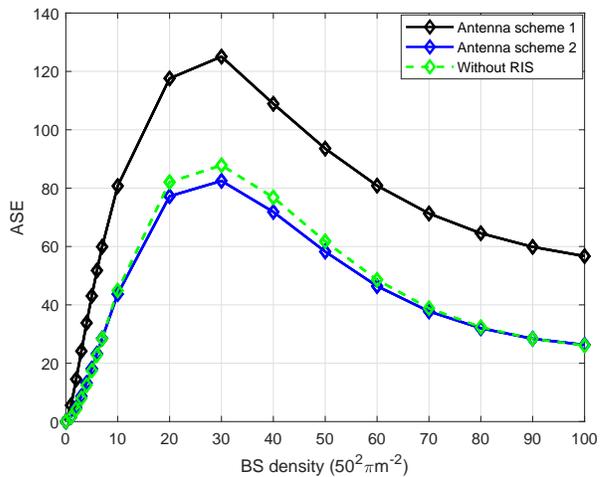}
  \caption{ASE as functions of BS density.}\label{ASE-BS-density}
\end{figure}

Fig. \ref{ASE-BS-density} shows the ASE as functions of the BS density. We can observe that the BS density exhibit a similar tendency with the coverage probability. At first, the ASE increases when the BS density is increased. Afterwards, the ASE decreases while its slope tends to be smaller. Note that the optimal BS density maximizing the ASE is larger than that maximizing the coverage probability. The reason is that although the network performance becomes worse in terms of the coverage probability, the performance loss can be compensated by the increasing BS density and the ASE still improves. In addition, the ASE of the RIS-assisted network is larger than that of the traditional networks without RISs, which demonstrates the benefit of deploying RISs in the cellular networks.

\begin{figure}
  \centering
  \includegraphics[width=3.5in]{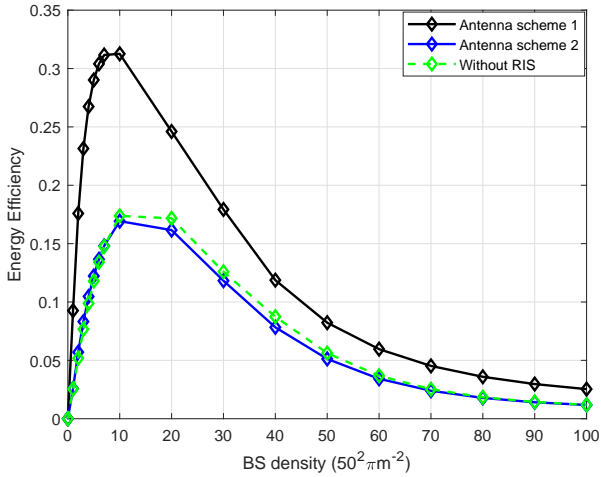}
  \caption{Energy efficiency as functions of BS density.}\label{EE-BS-density}
\end{figure}

Fig. \ref{EE-BS-density} illustrates the energy efficiency as functions of the BS density. We can observe that the tendency for the energy efficiency is similar with that for the coverage probability and the ASE. This can be explained as follows. When the BS density is relatively small, the energy efficiency increases since the coverage probability increases. However, when the BS density further increases, increasing power consumption and the decreasing coverage probability contributes to the decrement of the energy efficiency. In addition, the results show that introducing RISs into cellular networks can efficiently improve the energy efficiency of the dense cellular networks.

\begin{figure}
  \centering
  \includegraphics[width=3.5in]{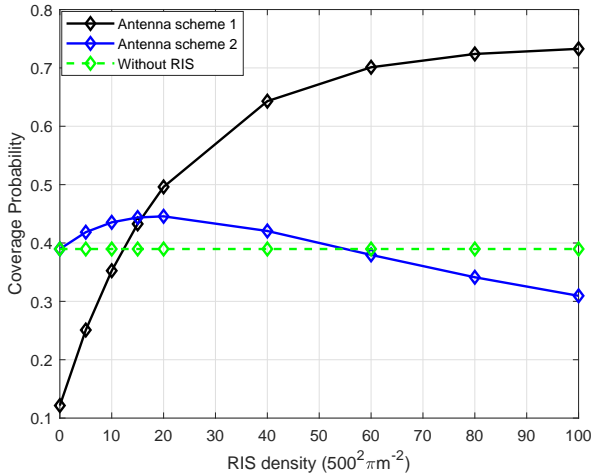}
  \caption{Coverage probability as functions of RIS density under the constant BS density.}\label{cp-RIS-density-constant-BS-density}
\end{figure}

Fig. \ref{cp-RIS-density-constant-BS-density} illustrates the coverage probability as functions of the RIS density under the constant BS density. It can be observed that $\mathcal{P}_1$ increases when increasing the RIS density. When the RIS density is relatively small, $\mathcal{P}_2$ shows a significant advantage over $\mathcal{P}_1$, which is even worse than $\mathcal{P}_t$. However, $\mathcal{P}_1$ increases significantly when increasing the RIS density while $\mathcal{P}_2$ goes gently below $\mathcal{P}_t$. The results indicate that deploying RISs is not always beneficial for the traditional networks. In order to obtain the optimal coverage probability, the RIS density needs to be carefully selected under different antenna schemes. The reason can be stated as follows. At the beginning, the distance between $u_0$ and the assisting RIS is much larger than that between $u_0$ and its serving BS. Hence, better performance can be achieved when the main lobe is directed towards the assisting RIS. When the RIS density increases, the interference from the RISs increases remarkably and the enhancement of the signal strength under antenna scheme 2 is limited due to that the main energy is directed towards the serving BS.

\begin{figure}
  \centering
  \includegraphics[width=3.5in]{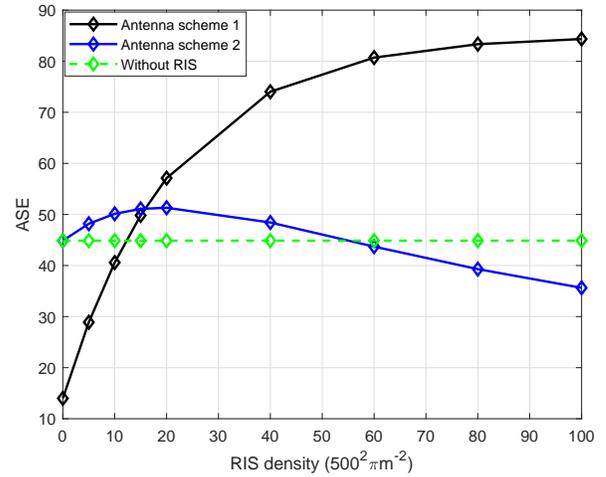}
  \caption{ASE as functions of RIS density under the constant BS density.}\label{ASE-RIS-density-constant-BS-density}
\end{figure}

Fig. \ref{ASE-RIS-density-constant-BS-density} illustrates the ASE as functions of the RIS density. We can observe that the ASE of the RIS-assisted networks under both antenna schemes, i.e., $\zeta_1$ and $\zeta_2$, exhibit the similar trends with the coverage probability. Note that the ASE is related to two components, i.e., the density of the active links and the coverage probability for a single link. When the RIS density is small, a fraction of users is served by both the direct links and assisting links while other users are served by the direct links. As the RIS density increases, all the users will receive the service of the assisting RISs. Hence, the ASE initially improves with the increasing RIS density. In the scenario where the RIS density is much larger than the BS density, the number of active links is limited by the BS density. Meanwhile, $\mathcal{P}_2$ decreases due to increasing interference from the RISs. Therefore, $\zeta_2$ degrades in the large RIS density regime.

\begin{figure}
  \centering
  \includegraphics[width=3.5in]{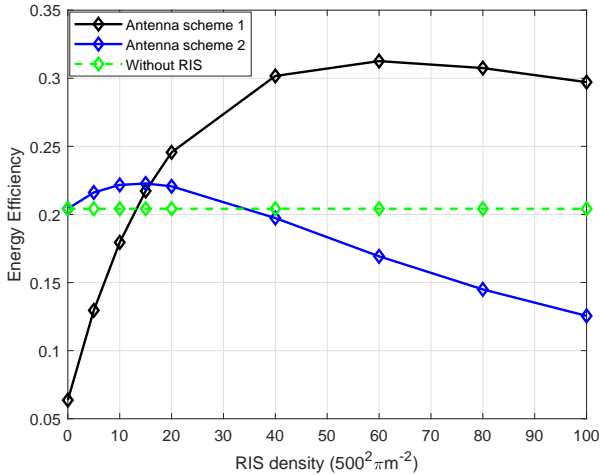}
  \caption{Energy efficiency as functions of RIS density under the constant BS density.}\label{EE-RIS-density-constant-BS-density}
\end{figure}

Fig. \ref{EE-RIS-density-constant-BS-density} plots the energy efficiency as functions of the RIS density. Note that the energy efficiency under antenna scheme 2 increases at first and then decreases rapidly when increasing the RIS density. The reason is that the increasing power consumption cannot be compensated by a slight increment in the coverage probability. In contrast, the energy efficiency under antenna scheme 1 improves when increasing the RIS density and then becomes flat when the RIS density is sufficiently large, which indicates that introducing more RISs is always energy efficient under antenna scheme 1.

\begin{figure}
  \centering
  \includegraphics[width=3.5in]{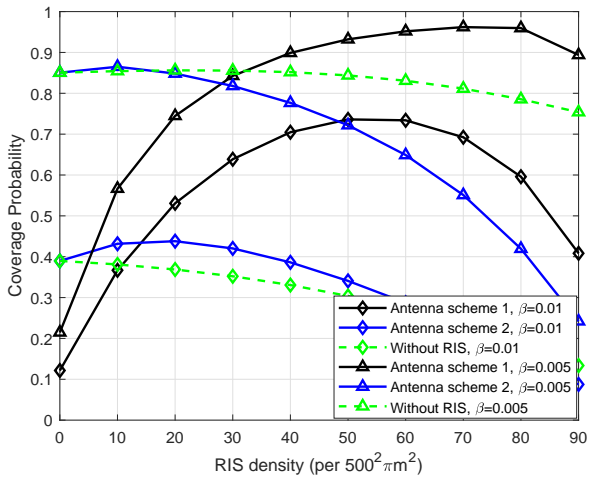}
  \caption{Coverage probability as functions of RIS density under the variable BS density.}\label{cp-RIS-density-variable-BS-density}
\end{figure}

In Fig. \ref{cp-RIS-density-variable-BS-density}, we investigate the tradeoff between the BS and RIS densities. In this simulation, the BS density is reduced by 1 per $500^2\pi$ when the RIS density is increased by 10 per $500^2\pi$. Since the BS density gradually decreases, the coverage probability of the traditional networks degrades accordingly. Different from the constant BS density case, there exists an optimal BS density $\lambda_{BS}^*$ maximizing the coverage probability under both antenna schemes while the optimal RIS density of $\mathcal{P}_1$ is larger than that of $\mathcal{P}_2$. Note that the coverage probabilities under both antenna schemes increase with $\left|\lambda_{BS}-\lambda_{BS}^*\right|$. Moreover, the performance gap between $\mathcal{P}_1$ and $\mathcal{P}_t$ when the blockage parameter $\beta=0.005$ is smaller than that when $\beta=0.01$, which indicates that the mmWave networks with dense blockages, i.e., the downtown areas of large cities, will benefit more from deploying RISs in terms of coverage probability.

\begin{figure}
  \centering
  \includegraphics[width=3.5in]{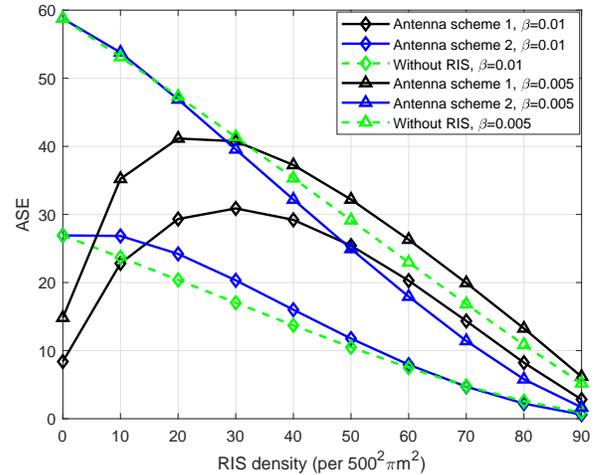}
  \caption{ASE as functions of RIS density under the variable BS density.}\label{ASE-RIS-density-variable-BS-density}
\end{figure}

Fig. \ref{ASE-RIS-density-variable-BS-density} illustrates the ASE as functions of the RIS density. We can observe that $\zeta_2$ and $\zeta_t$ decreases when the RIS density increases and the BS density decreases. $\zeta_1$ exhibits a similar trend to $\mathcal{P}_1$. The reason can be explained as follows. When the RIS density decreases, the reduction in the coverage probability caused by the increasing interference cannot be compensated by the signal strength enhancement due to that the main lobe is directed towards the BS. Similar to the coverage probability, the performance gain of $\zeta_1$ over $\zeta_t$ when $\beta=0.01$ is larger than that when $\beta=0.005$.

\begin{figure}
  \centering
  \includegraphics[width=3.5in]{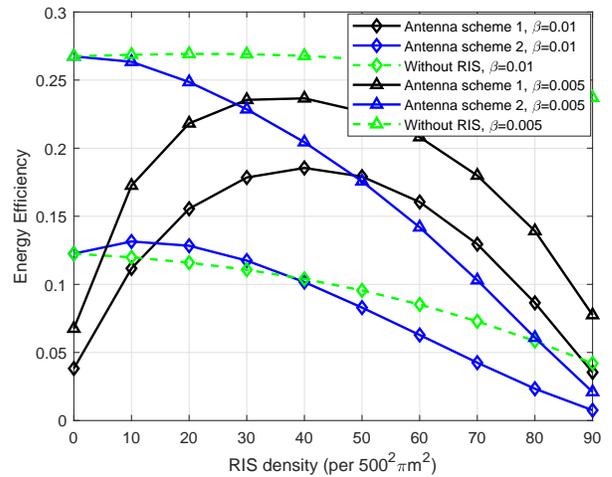}
  \caption{Energy efficiency as functions of RIS density under the variable BS density.}\label{EE-RIS-density-variable-BS-density}
\end{figure}

Fig. \ref{EE-RIS-density-variable-BS-density} illustrates the energy efficiency as functions of the RIS density. It can be observed that the optimal RIS density maximizing the energy efficiency is smaller than that optimizing the coverage probability. The reason is that the increasing power consumption cannot be compensated by the increment in the coverage probability. In addition, the RIS-assisted networks show a significant advantage over the traditional cellular networks in terms of energy efficiency when $\beta=0.01$ while deploying RISs causes the degradation in the energy efficiency when $\beta=0.005$, which indicates that the mmWave networks with dense blockages will benefit more from deploying RISs in terms of energy efficiency.

\begin{figure}
  \centering
  \includegraphics[width=3.5in]{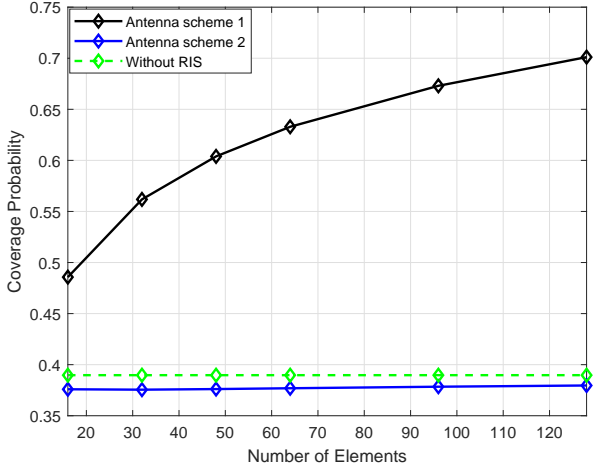}
  \caption{Coverage probability as functions of number of elements.}\label{cp-RIS-size}
\end{figure}

Fig. \ref{cp-RIS-size} illustrates the ASE as functions of the RIS size. We can observe that $\mathcal{P}_1$ increases with the RIS size. The reason is that the signal strength is enhanced with the increasing RIS size.

\section{Conclusion}
In this paper, we have investigated the coverage probability and energy efficiency in a RIS-assisted mmWave network. We have derived the expressions of the coverage probability and the area spectral efficiency. The coverage probability under the special case where the blockage parameter is sufficiently small has been also derived. Numerical results have demonstrated that better coverage performance and higher energy efficiency can be achieved by the large-scale deployment of RISs. In addition, the tradeoff between the BS and RIS densities is investigated and the results show that the RISs are excellent supplementary for the traditional networks to improve the coverage probability with limited power consumption. In particular, the mmWave networks with dense blockages will benefit more from deploying RISs in terms of coverage probability and energy efficiency.

\section{Appendices}
\subsection{Proof of Lemma \ref{lemma-1}}
For $\rho=\{L,N\}$, let $x_{d,\rho}$ be the distance between $u_0$ and its nearest BS in $\Phi_{BS,\rho}$. $u_0$ is associated with a BS in $\Phi_{BS,L}$ if and only if it has an LOS BS and the path loss of the nearest BS in $\Phi_{BS,L}$ is smaller than that of the nearest BS in $\Phi_{BS,N}$. Hence, the probability that $u_0$ is associated with a LOS BS is derived as
\begin{equation}
\begin{split}
A_{d,L}&=\mathbbm{P}\left[C_Lx_{d,L}^{-\alpha_L}>C_N\nu_{d,N}^{-\alpha_N}\right]\\
&\overset{(a)}{=}\int_{0}^{\infty}\mathbbm{P}\left[x_{d,N}>\chi_{L}(r)\right]f_{d,L}(r)\text{d}r,
\end{split}
\end{equation}
where (a) follows from the PDF of $\nu_{d,L}$.
\begin{equation}
\begin{split}
\mathbbm{P}\left[x_{d,N}>\chi_{L}(x)\right]&=\mathbbm{P}\left[\Phi_{BS,N}\cap\mathcal{B}(0,\chi_{d,L}(r)=\emptyset)\right]\\
&=e^{-2\pi\lambda_{BS}\int_{0}^{\chi_{d,L}}(1-p(r))r\text{d}r}.
\end{split}
\end{equation}

\subsection{Proof of Lemma \ref{lemma-2}}
$u_0$ is associated with a assisting RIS when two conditions are satisfied. The first condition is that $u_0$ is on the RIS-equipped side of the blockage. The second condition is that $u_0$ and its serving BS are on the same side of the RIS. Note that the probability that the first condition is met is $\frac{1}{2}$.

Next, we derive the probability that the second condition is satisfied. Note that the user and the BS are located on the same side of the RIS when $\upsilon_u\geq\upsilon_g$. Hence, we need to derive the probability is equal to the probability that $\upsilon_u$ is larger than or equal to $\upsilon_g$. Let $z$, $\upsilon_{u}$ and $\upsilon_{g}$ be the distance between the serving BS and the assisting RIS, the angle between the user-RIS link and the RIS, the angle between the user-RIS link and the BS-RIS link, respectively. Then $z$ can be obtained as follows
\begin{equation}
z^2=x^2+y^2-2xy\cos(\upsilon).
\end{equation}

Similarly, $\upsilon_{g}$ can be expressed as a function of $x$, $y$ and $\upsilon$, which is shown as
\begin{equation}
\begin{split}
&x^2=y^2+z^2-2yz\cos(\upsilon_g)\\
&\overset{(a)}{=}y^2+x^2+y^2-2xy\cos(\upsilon)\\
&-2y\sqrt{x^2+y^2-2xy\cos(\upsilon)}\cos(\upsilon_g).
\end{split}
\end{equation}

Therefore, $\upsilon_g$ can be expressed as
\begin{equation}
\upsilon_g=\cos^{-1}\frac{y-x\cos(\upsilon)}{\sqrt{x^2+y^2-3xy\cos(\upsilon)}}.
\end{equation}

The probability that the second condition is met can be derived as
\begin{equation}
\begin{split}
\mathcal{C}(x,y,\upsilon)&=\mathbbm{P}\left[\upsilon_u\geq\upsilon_g\right]\\
&=\mathbbm{P}\left[\upsilon_u\geq\cos^{-1}\left(\frac{y-x\cos(\upsilon_u)}{\sqrt{x^2+y^2-2xy\cos(\upsilon_u)}}\right)\right]\\
&=\overset{(a)}{=}1-\frac{1}{\pi}\cos^{-1}\left(\frac{y-x\cos(\upsilon_u)}{\sqrt{x^2+y^2-2xy\cos(\upsilon_u)}}\right),
\end{split}
\end{equation}
where (a) follows from $\upsilon_u\sim\text{U}(0,\pi)$.

\subsection{Proof of Theorem \ref{theorem-1}}
The coverage probability can be derived as follows
\begin{equation}\label{STP-derivation}
\begin{split}
&\mathbbm{P}(\text{SINR}_{\rho}>T)\\
=&\mathbbm{P}\left(\frac{\left|\left(\sqrt{PL_{d,0}M_{\text{DL},0}}+\sqrt{PL_{g,0}L_{u,0}M_{\text{RL},0}}\right)\right|^2}
{I_{BS}+I_{R}+I_{R,\text{idle}}+\sigma^2}>T\right)\\
=&\mathbbm{P}\left(1>\frac{T\left(I_{BS}+I_{R}+I_{R,\text{idle}}+\sigma^2\right)}{\left|\left(\sqrt{PL_{d,0}M_{\text{DL},0}}+\sqrt{PL_{g,0}L_{u,0}M_{\text{RL},0}}\right)\right|^2}
\right)\\
\overset{(a)}{\approx}&\mathbbm{P}\left(\nu>\frac{T\left(I_{BS}+I_{R}+I_{R,\text{idle}}+\sigma^2\right)}{\left|\left(\sqrt{PL_{d,0}M_{\text{DL},0}}+\sqrt{PL_{g,0}L_{u,0}M_{\text{RL},0}}\right)\right|^2}
\right)\\
\overset{(b)}{\approx}&1-\mathbbm{E}_{x,y,\theta}\left[1-e^{-s_{\rho}\sigma^{2}}
\mathcal{L}_{I_{BS}}(s_{\rho})\mathcal{L}_{I_R}(s_{\rho})\mathcal{L}_{I_{R,\text{idle}}}(s_{\rho})\right]^{W}\\
\overset{(c)}{=}&\mathbbm{E}_{x,y,\theta}\left[\sum_{\rho\{L,N\}}\sum\limits_{w=1}^{W}\binom{W}{w}(-1)^{w+1}e^{-s_{\rho}\sigma^{2}}
\mathcal{L}_{I_{BS,L}}(s_{\rho})\right.\\
&\left.\mathcal{L}_{I_{BS,N}}(s_{\rho})\mathcal{L}_{I_{R,L}}(s_{\rho})\mathcal{L}_{I_{R,N}}(s_{\rho})\mathcal{L}_{I_{R,L,\text{idle}}}(s_{\rho})
\mathcal{L}_{I_{R,N,\text{idle}}}(s_{\rho})\right]\\
\overset{(d)}{=}&\frac{1}{2\pi}\sum_{\rho\{L,N\}}\int_{0}^{2\pi}\int_{0}^{\infty}\int_{0}^{\infty}\sum\limits_{w=1}^{W}\binom{W}{w}(-1)^{w+1}e^{-s_{\rho}\sigma^{2}}\\
&\mathcal{L}_{I_{BS,L}}(s_{\rho})\mathcal{L}_{I_{BS,N}}(s_{\rho})\mathcal{L}_{I_{R,L}}(s_{\rho})\mathcal{L}_{I_{R,N}}(s_{\rho})\\
&f_{d,\rho}(x)f_{u,\rho}(y)\text{d}x\text{d}y.
\end{split}
\end{equation}

In (a), we utilize a ``dummy'' gamma variable $\nu$ with shape parameter $W$ and unit mean to approximate the constant number one and $\nu$ converges to one when $W$ geoes to infinity, i.e., $\lim_{w\rightarrow\infty}\frac{w^wx^{w-1}e^{-wx}}{\Gamma(w)}=\delta(x-1)$ \cite{gamma}, where $\delta(x)$ is the Dirac delta function. (b) is from Alzer's inequality \cite{Alzer}. (c) follows from the binomial theorem and the independence between $I_{BS,L}$, $I_{BS,N}$, $I_{R,L}$ and $I_{R,N}$. (d) follows from the PDF of the distance between $u_0$ and its serving BS/RIS.

The Laplace transform of the interference from the LOS BSs $\Phi_{BS,L}$ can be derived as follows
\begin{equation}
\begin{split}
\mathcal{L}_{I_{BS,L}}(s_{\rho})&=\mathbbm{E}_{I_{BS,L}}\left[e^{\sum_{i\in\tilde{\Phi}_{BS,L}}s_{\rho}P\bar{m}_{BS,\text{DL}}
\bar{m}_{u,\text{DL}}L_{d,i}}\right]\\
&=\mathbbm{E}_{I_{BS,L}}\left[\prod_{i\in\tilde{\Phi}_{BS,L}}e^{s_{\rho}P\bar{m}_{BS,\text{DL}}
\bar{m}_{u,\text{DL}}L_{d,i}}\right]\\
&\overset{(a)}{=}\exp\left(-2\pi\lambda_{BS}p_{a,BS}\right.\\
&\left.\int_{\chi(x)}^{\infty}e^{-\beta r}\left(1-e^{-s_{\rho}P\bar{m}_{BS}\bar{m}_uC_Lr^{-\alpha_L}}\right)r\text{d}r\right).
\end{split}
\end{equation}
where (a) follows from the probability generating functional (PGFL) of PPP. Note that the Laplace transform of the interference from the NLOS BSs can be derived following the similar steps.

The Laplace transform of the interference from the LOS RISs $\Phi_{R,L}$ can be derived as follows
\begin{equation}
\begin{split}
&\mathcal{L}_{I_{R,L}}(s_{\rho})\\
&=\mathbbm{E}_{I_{R,L}}\left[e^{\sum_{i\in\tilde{\Phi}_{R,L}}s_{\rho}\bar{m}_{r,\text{RL}}\bar{m}_{BS,\text{RL}}
\bar{m}_{u,\text{RL}}\sum_{m_i\in\bar{\Phi}_{BS}}L_{g,m_i}}\right]\\
&\overset{(a)}{=}\mathbbm{E}_{I_{R,L}}\left[\exp\left(\sum_{i\in\tilde{\Phi}_{R,L}}s_{\rho}\bar{m}_{r,\text{RL}}\bar{m}_{BS,\text{RL}}
\bar{m}_{u,\text{RL}}\pi\lambda_{BS}\right.\right.\\
&\left.\left.\int_{0}^{\infty}\left(C_Lz^{-\alpha_L}e^{-\beta z}+C_Nz^{-\alpha_N}(1-e^{-\beta z})\right)\text{d}z\right)\right]\\
&=\exp\left(-\pi\lambda_Rp_{a,R}\int_{y}^{\infty}e^{-\beta r}\left(1-\exp\left(\sum_{i\in\tilde{\Phi}_{R,L}}s_{\rho}\bar{m}_{r,\text{RL}}\right.\right.\right.\\
&\left.\left.\left.\bar{m}_{BS,\text{RL}}
\bar{m}_{u,\text{RL}}\pi\lambda_{BS}\int_{0}^{\infty}\left(C_Lz^{-\alpha_L}e^{-\beta z}+C_Nz^{-\alpha_N}\right.\right.\right.\right.\\
&\left.\left.\left.\left.(1-e^{-\beta z})\right)\text{d}z\right)\right)r\text{d}r\right),
\end{split}
\end{equation}
where (a) follows from Campbell's theorem.

\end{spacing}
\end{document}